\documentclass[10pt]{wlscirep}
\usepackage[utf8]{inputenc}
\usepackage[T1]{fontenc}
\usepackage{graphicx}
\title{\textbf{Particle Levitation Velocimetry for boundary layer measurements in high Reynolds number liquid helium turbulence}
}

\author[1,2]{Yinghe Qi}
\author[1,2,*]{Wei Guo}
\affil[1]{National High Magnetic Field Laboratory, 1800 East Paul Dirac Drive, Tallahassee, Florida 32310, USA}
\affil[2]{Mechanical Engineering Department, FAMU-FSU College of Engineering, Tallahassee, Florida 32310, USA}

\affil[*]{wguo@magnet.fsu.edu}

\begin{abstract}
Understanding boundary layer flows in high Reynolds number (Re) turbulence is crucial for advancing fluid dynamics in a wide range of applications, from improving aerodynamic efficiency in aviation to optimizing energy systems in industrial processes. However, generating such flows requires complex, power-intensive large-scale facilities. Furthermore, the use of local probes, such as hot wires and pressure sensors, often introduces disturbances due to the necessary support structures, compromising measurement accuracy. In this paper, we present a solution that leverages the vanishingly small viscosity of liquid helium to produce high Re flows, combined with an innovative Particle Levitation Velocimetry (PLV) system for precise flow-field measurements. This PLV system uses magnetically levitated superconducting micro-particles to measure the near-wall velocity field in liquid helium. Through comprehensive theoretical analysis, we demonstrate that the PLV system enables quantitative measurements of the velocity boundary layer over a wall unit range of $44\le y^{+}\le 4400$, with a spatial resolution that, depending on the particle size, can reach down to about 10~$\mu$m. This development opens new avenues for exploring turbulence structures and correlations within the thin boundary layer that would be otherwise difficult to achieve.
\end{abstract}
\begin{document}

\flushbottom
\maketitle
\thispagestyle{empty}

%\noindent Please note: Abbreviations should be introduced at the first mention in the main text – no abbreviations lists. Suggested structure of main text (not enforced) is provided below.

\section*{Introduction}
Turbulent boundary layer flows are fundamental to numerous engineering applications. For instance, dissipation in these flows controls aerodynamic forces and heating in hypersonic vehicles and accounts for over 50\%  of surface drag in aircraft, as well as nearly 95\% of energy loss in long-distance pipeline transport~\cite{Womack_Meneveau_Schultz_2019,10.1115/1.2817367}. Many of these applications involve flows at very high Reynolds numbers (Re), where the turbulent boundary layer is characterized by near-wall small-scale structures, such as streaks and vortices, as well as larger, energetic eddies in the outer region~\cite{Smits-review-2011,jimenez2013near}. The interaction between these scales at high Re intensifies turbulence and increases the complexity of flow dynamics. To better understand these interactions, it is crucial to precisely quantify turbulence characteristics within the boundary layer, including the scaling behaviors of mean velocity and turbulence intensity. This knowledge is vital for developing more accurate and predictive turbulence models, which are essential for optimizing designs and enhancing performance in practical engineering applications.

The mean velocity profile $\overline{U}_x(y)$ in the streamwise direction near a solid wall in fully developed turbulent flows has been studied extensively and is known to consist of three distinct regions along the coordinate \emph{y} perpendicular to the wall~\cite{Smits-review-2011,jimenez2013near,white2006viscous}. The inner region, dominated by viscous effects, typically extends from the wall to $y^+=y/y^*=50$, where $y^*= \nu/u_\tau$ denotes the viscous length scale~\cite{patel2016influence}. In this expression, $\nu$ is the kinetic viscosity, and $u_\tau=(\tau_w/\rho_f)^{1/2}$ is the viscous velocity, with $\tau_w$ being the wall shear stress and $\rho_f$ the fluid density. At sufficiently large $y$, for example beyond $y/R \approx 0.12$ in pipe flows with a pipe radius $R$, the wake region emerges, where $\overline{U}_x(y)$ depends on overall flow conditions. Between these two regions lies the overlap region, where $\overline{U}_x(y)$ follows a universal logarithmic profile, commonly referred to as the ``law of the wall'' or ``log law'', expressed as \(U^+=\frac{1}{\kappa}\ln{y^+}+B\)~\cite{Smits-review-2011,jimenez2013near,white2006viscous}. Here, $U^+=\overline{U}_x/u_\tau$, and $\kappa$ and \emph{B} are the von K\'arm\'an constant and the additive constant, respectively. Despite extensive experimental\cite{PhysRevLett.108.094501,furuichi2015friction, zagarola1996experiments, osterlund2000note, orlu2017reynolds, monty2005developments} and numerical \cite{wu2008direct, lee2015direct, yamamoto2018numerical}investigations, some key issues remain unsolved, such as the extent and Re dependence of the log law and the precise values of these constants~\cite{marusic2010wall,mckeon2007introduction}. The Princeton Superpipe experiments suggest that the log law appears in the range $600\leqslant y^+ \leqslant0.12 R/y^*$ when the pipe Reynolds number exceeds about $2.3\times10^5$~\cite{Smits-review-2011, zagarola1998mean,mckeon2004further,PhysRevLett.108.094501}. These experiments reported a von K\'arm\'an constant $\kappa=0.42$, which differs from the typical values of 0.37–0.39 observed in high-Re boundary layer and channel flows \cite{nagib2007approach,osterlund2000note,monty2005developments,zanoun2003evaluating,nickels2007some}, raising questions about the universality of $\kappa$ across different flow types. However, more recent high-Re pipe flow data from Furuichi \emph{et al.}, using the ``Hi-Reff'' facility in Japan, suggest a $\kappa$ value of 0.385, indicating a potential degree of universality for $\kappa$~\cite{furuichi2015friction,furuichi2018further}. Given the pivotal role of $\kappa$ in modeling and numerical simulations of wall-bounded flows, further independent high-Re flow measurements are necessary.

In high-Re turbulence experiments, velocity field measurements have predominantly relied on hot-wire anemometers and pressure sensors~\cite{stainback1993review,vukoslavcevic1991velocity,wilmarth1975pressure}. These sensors normally have relatively large sizes, limiting their spatial resolution and effectiveness in capturing the fine details of boundary layer flows. For instance, conventional hot-wire sensors are difficult to reduce below 0.25 mm in length. In response, efforts have been made to miniaturize these sensors. As an example, Smits' team developed hot-wire sensors with an active length as small as 30~$\mu$m~\cite{bailey2010turbulence,PhysRevLett.108.094501}. Similarly, advances have also been made in pressure sensor technology~\cite{javed2019review}. However, despite these miniaturization efforts, both types of sensors still face one challenging issue: they require support structures attached to the wall, which inevitably introduce flow disturbances. These disturbances compromise the collection of clean data in high-Re turbulent flows, particularly in thin boundary layers, where minimizing interference is crucial for accurate measurements. In addition to hot-wire anemometry and pressure sensors, non-intrusive measurement techniques based on direct flow visualization have been widely applied to measure velocity fields in various types of flows. Particle Image Velocimetry (PIV) and Particle Tracking Velocimetry (PTV) are two commonly used flow visualization methods~\cite{adrian1991particle,van2013piv,de2014high, scarano2012tomographic,adamczyk19882,dabiri2019particle, Mastracci-2018,Mastracci-2019,Tang-2020,Tang-2023}. However, both methods face limitations in spatial resolution when applied to velocity field measurements in high-Re turbulent boundary layers, restricting their ability to resolve thin boundary layer structures~\cite{lavoie2007spatial,hoyer20053d}. Another non-intrusive technique is Molecular Tagging Velocimetry (MTV), which has been applied in various flow conditions, including studies of wall shear stress and velocity profiles~\cite{li2021review,miles1991turbulent,michael2011femtosecond,hsu2009two,Varga-2018,Gao-2015,Marakov-2015}. However, typical MTV setups can only capture the velocity component perpendicular to the tracer lines, and their spatial resolution is constrained by the displacement of the tracer lines, typically on the order of $10^2$ $\mu$m \cite{guo2019molecular}. This limitation reduces its effectiveness for capturing fine-scale boundary layer structures in high-Re turbulence.

To overcome these limitations, we propose an approach that combines liquid helium (LHe) as the working fluid with a novel measurement technique called Particle Levitation Velocimetry (PLV), which utilizes magnetically levitated superconducting micro-particles as probes. LHe's exceptionally low kinematic viscosity, i.e., nearly three orders of magnitude lower than that of ambient air~\cite{Donnelly-1998-JPCRD}, allows for the generation of very high-Re flows in compact facilities~\cite{donnelly2012high,skrbek1999turbulent,donnelly2012flow}, which is difficult to achieve with conventional fluids. Moreover, LHe provides a cryogenic environment suitable for superconducting coils and particles, enabling magnetic levitation of micro-particle probes in LHe without physical supports to disturb the flow.

In this paper, we present the design and analysis of the PLV system, integrated with our Liquid Helium Flow Visualization Facility (LHFVF), which can generate turbulent pipe flows with Re exceeding $10^6$~\cite{sanavandi2020cryogenic}. The PLV system uses a compact four-coil setup to create a three-dimensional trap with adjustable size and streamwise gradient, enabling easy particle loading while stably levitating superconducting micro-particles near the pipe wall, preventing them from being swept away by high-Re flows. Comprehensive simulations were conducted to calculate the potential energy of the particles in the magnetic trap and to model their motion under both static and high-Re flow conditions. The results show that when the flow is initiated, the particles move downstream due to the fluid's drag force, undergoing damped oscillations before settling into a new equilibrium position. By measuring the particle’s mean displacement, we can determine the drag force and consequently calculate the mean flow velocity at the corresponding levitation height. Furthermore, by introducing random velocity fluctuations into the simulations, we established the correlation between particle position fluctuations and velocity fluctuations in the flow, enabling the evaluation of near-wall velocity fluctuations. The levitation height can be adjusted continuously by tuning the coil currents. Our simulations indicate that the PLV system can measure the velocity field across a wide range of wall units $44\le y^{+} \le 4400$. By implementing multiple levitated particles, this system may provide new opportunities to study turbulence structures and correlations in the boundary layer that would be otherwise difficult to achieve.

\section*{Results}
\subsection*{Liquid Helium Flow Facility and Particle Levitation}
\begin{figure}[ht]
\centering
\includegraphics[width=1.0\linewidth]{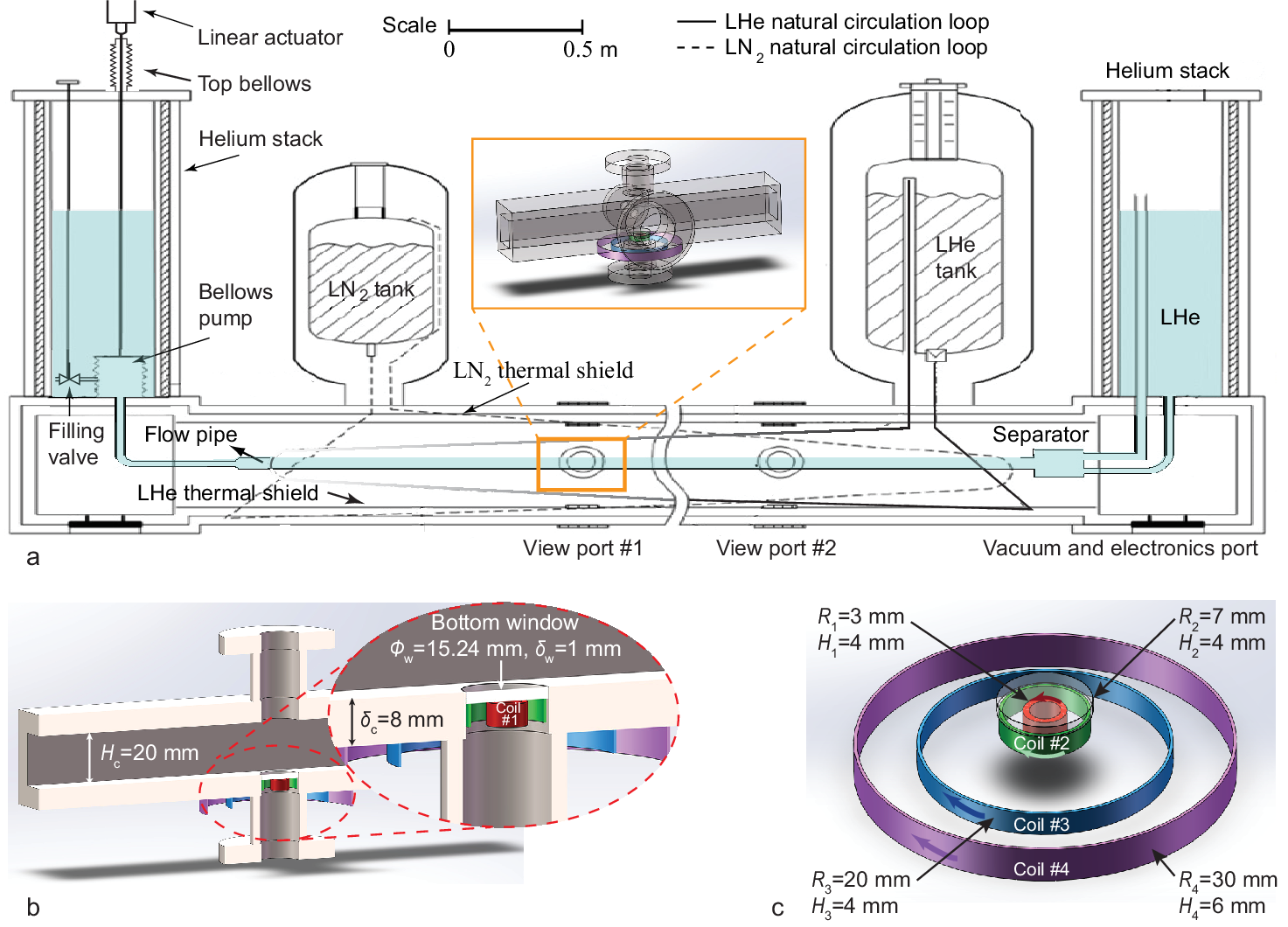}
\caption{\textbf{Pipe flow and particle levitation facility. a} Schematic diagram of the Liquid Helium Flow Visualization Facility. \textbf{b} A schematic showing the locations of the coils for particle levitation. \textbf{c} A schematic showing the coil specifications.}
\label{fig:Figure 1}
\end{figure}

As shown schematically in Figure~\ref{fig:Figure 1}a, the LHFVF is a cryostat designed for generating and visualizing LHe pipe flows. This facility includes a 5-meter-long horizontal cylindrical chamber with an inner diameter of 0.2~m, housed inside the evacuated cryostat body and enclosed by two concentric radiation shields that are cooled by natural circulation loops connected to the liquid helium and nitrogen tanks. A 3.35-meter-long pipe with a square cross-section ($2\times2$ cm$^2$) is positioned at the chamber center, connected to helium storage stacks at both ends, where the LHe temperature is controlled by regulating the vapor pressure. The system is equipped with three sets of windows: two vertical sets to allow laser beams to pass through the top and bottom of the pipe, and a side window set for capturing flow images using a high-speed camera. Each set consists of three windows mounted on the vacuum and the two radiation shields. To drive the LHe flow, a bellows pump is installed in the left helium storage stack. This bellows pump, with a cross-sectional area of $1.8\times10^{-2}$ m$^2$ and a stroke length of 9.4 cm, can displace up to 1.7 liters of LHe. The bellows is connected to a linear actuator, driven by a computer-controlled stepper motor, capable of producing enough thrust to drive the LHe through the pipe at pipe Reynolds numbers exceeding $10^6$~\cite{sanavandi2020cryogenic}.

To probe the flow, we propose to adopt the PLV system, which consists of superconducting niobium particles, i.e., micro-spheres with diameters $d_p=10-50$ $\mu$m, and a superconducting coil system for levitating them. Below its critical temperature $T_c=9.2$~K, niobium becomes superconducting. At LHe temperature of 4.2~K, the lower critical field of niobium is about 0.15~T~\cite{ikushima1969superconductivity}. Below this field strength, the niobium particles exhibits perfect diamagnetism, with a volume magnetic susceptibility $\chi$ close to -1 as compared to typical values of -10$^{-5}$ to -10$^{-6}$ for ordinary diamagnetic materials \cite{yamato2020magnetic,finnemore1966superconducting}. This makes them much easier to levitate using relatively weak magnetic fields. Additionally, niobium is easier to fabricate and machine than many other superconducting materials, making it a practical choice for used in the PLV system.

When a superconducting niobium particle is placed in a magnetic field $\mathbf{B}(\mathbf{r})$, it experiences a potential energy per unit volume given by \cite{arrayas2021design}:
\begin{equation}
E(\emph{\textbf{r}})=\left(\rho_p-\rho_f\right) g y+\frac{1}{4} \frac{\textbf{B}^2(\emph{\textbf{r}})}{\mu_0}
\label{eq:energy}
\end{equation}
where $\rho_p=8570$ kg/m$^3$ and $\rho_f=145$ kg/m$^3$ are the densities of niobium and LHe, respectively, $g$ represents gravitational acceleration, and $\mu_0$ is the vacuum permeability. Levitation of the niobium particle can be achieved at a location where $\partial E(\mathbf{r})/\partial z=0$. For stable levitation, $E(\mathbf{r})$ must increase in all directions from the levitation location.

Given the LHe environment, we choose to use superconducting coils to generate the magnetic field to avoid Joule heating. There are various design constraints to consider: the coils must fit around the window flange of the flow pipe, and the current in the coil wires must remain below the critical current to maintain the superconducting state. At the same time, the coils must provide a strong magnetic field gradient in both the vertical and the flow directions to levitate the niobium particles and to prevent the particles from being swept away by LHe in high Re flows. After thoroughly evaluating various coil configurations and current settings, we finally arrived at the optimal coil design, as illustrated schematically in Figure~\ref{fig:Figure 1}b and \ref{fig:Figure 1}c.

This coil system consists of four concentric coils, coaxially aligned with the bottom window. Coils \#1 and \#2 are installed inside the window flange just below the 1-mm-thick bottom window, while Coils \#3 and \#4 are placed outside the window flange beneath the flow pipe. The radii of Coils \#1, \#2, \#3, and \#4 are 3 mm, 7 mm, 20 mm, and 30 mm, respectively. Each coil is constructed using copper-niobium titanium wires from SUPERCON Inc, capable of carrying up to $I_c=6.8$~A at 4.2~K\cite{hemmati2009drag}. Coils \#1, \#2, and \#3 each consist of 15 layers with 40 turns per layer, yielding a total of $N_1=N_2=N_3=600$ turns per coil. With a wire thickness of 102~$\mu$m, 40 turns in each layer results in a coil height of 4 mm. Coils \#4 also has 15 layers but 60 turns per layer, rendering a total of $N_4=900$ turns and a coil height of 6 mm. As we will present in subsequent sections, when suitable currents are applied to these coils, a three-dimensional potential trap can form where $E(\mathbf{r})$ exhibits a local minimum at the levitation location. As indicated in Figure~\ref{fig:Figure 1}c, the current in coil \#1 flows in opposite direction to that in coils \#2, \#3, and \#4. This configuration lowers the potential energy at the trap center while raising it along the sides, creating pancake-shaped trap elongated in the horizontal direction. This design facilitates easy particle loading and allows for convenient particle displacement control within the trap region, accommodating a wide range of LHe flow speeds.

\subsection*{Particle loading and levitation}
\subsubsection*{Particle loading}
The niobium particles need to be placed inside the flow pipe before the cryostat is cooled down. During the cooling process, the pipe will be pumped and flushed with helium gas. When LHe starts filling, the liquid may slosh in the pipe. The magnetic trap can be turned on only after the pipe is fully filled with LHe and the temperature drops below the superconducting transition temperature of the coils. Without the magnetic trap active, the flowing gaseous and liquid helium may carry the particles away from the window region, preventing their levitation in later measurements. To avoid this, a small pit (diameter: 0.5~mm; depth: 0.25~mm) can be carved into the bottom window to securely contain the particles. The pit is located at 2.25 mm downstream from the center of the bottom window. As will be discussed later, the maximum streamwise displacement of the particles in all the considered LHe flows is less than 2 mm from the trap center. Therefore, the pit would not affect the particles during the boundary layer measurements.

\begin{figure}[ht]
\centering
\includegraphics[width=0.5\linewidth]{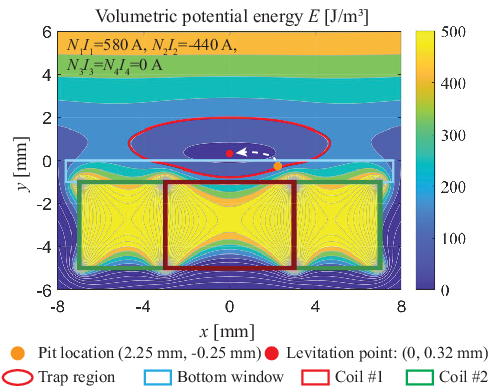}
\caption{\textbf{Contour plot of the volumetric potential energy for particle loading.} Only Coils \#1 and \#2 are activated.}
\label{fig:Figure 2}
\end{figure}

After the pipe is filled with LHe and cooled, we can transfer the niobium particles from the downstream pit to the center of the magnetic trap. To accomplish this, we active only Coils \#1 and \#2 by applying currents of $I_1=0.97$~A and $I_2=-0.73$~A, yielding $N_1I_1=580$~A and $N_2I_2=-440$~A. These currents are well below the critical current $I_c$ of the wire. The method for calculating the magnetic field $\mathbf{B}(\mathbf{r})$ generated by the coils is described in the Method section. Using the computed $\mathbf{B}(\mathbf{r})$, we can produce a contour plot of the volumetric potential energy $E(\mathbf{r})$, as shown in Figure~\ref{fig:Figure 2}. Given the concentric coil arrangement, $E(\mathbf{r})$ exhibits axial symmetry around the coil center axis. Figure~\ref{fig:Figure 2} provides a cross-section view in the $x-y$ plane, where x is the axis in the flow direction, and the coordinate origin is set at the center of the bottom window surface. As depicted, a region with closed contours of $E(\mathbf{r})$ is formed, where $E(\mathbf{r})$ decreases towards the region center. This is the trapping region, where a niobium particle always experiences a net force directing it towards the center, allowing it to be stably levitated. For clarity, the boundary of the trapping region is marked in red. When only Coils \#1 and \#2 are activated with the specified currents, the radial confinement is relatively weak, and the trapping region extends to a large radius, covering the pit used for niobium particle storage. In this case, a niobium particle initially placed in the pit (the orange dot in Figure~\ref{fig:Figure 2}) will experience a net lifting force that transports it to the trap center located at ($x_0=0$, $y_0=0.32$~mm), marked by the red dot.

\subsubsection*{Particle levitation and position control}
\begin{figure}[ht]
\centering
\includegraphics[width=\linewidth]{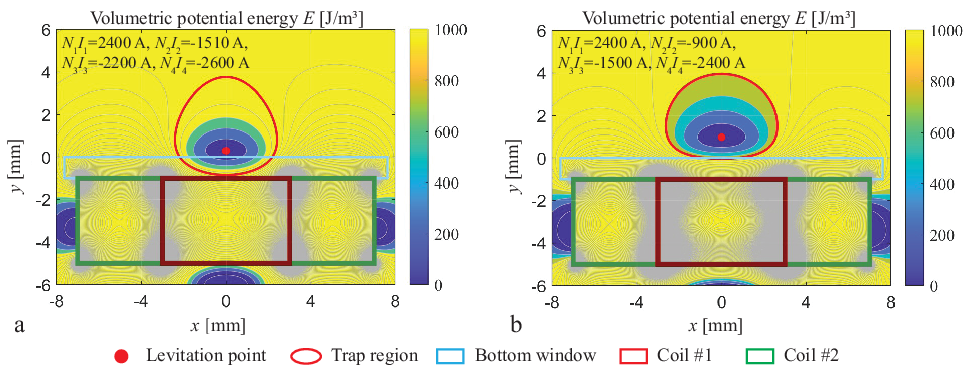}
\caption{\textbf{Contour plots of representative volumetric potential energy fields for particle levitation. a} Configuration with the particle levitated at $y_0=0.30$~mm. \textbf{b} Configuration with the particle levitated at $y_0=0.98$~mm }
\label{fig:Figure 3}
\end{figure}

After the niobium particle is loaded into the center of the magnetic trap, we can activate all four coils to create a more compact trapping region, suitable for flow field measurements in high-Re LHe flows. Figure~\ref{fig:Figure 3}a shows a representative contour plot of $E(\mathbf{r})$ with currents $I_1=4.00$~A, $I_2=-2.52$~A, $I_3=-3.67$~A, $I_4=-2.89$~A, yielding $N_1I_1=2400$~A, $N_2I_2=-1510$~A, $N_3I_3=-2200$~A, $N_4I_4=-2600$~A. To provide a clearer view of the $E(\mathbf{r})$ profile around the levitation point, Figure~\ref{fig:Figure 3} only shows the region near the bottom window and omits the visual representation of coils \#3 and \#4, although their contribution to $E(\mathbf{r})$ is fully accounted for in the calculations. In this configuration, the particle is levitated at $y_0=0.30$~mm. Compared to Figure~\ref{fig:Figure 2} where only Coils \#1 and \#2 are activated, the addition of Coils \#3 and \#4 increases $E(\mathbf{r})$ at larger |$x$| values. This results in a much more compact trapping region with steeper horizontal gradients. As we will discuss in the next section, the viscous drag force on the niobium particle from flowing LHe scales with $d_p^2$, while the magnetic restoring force, being a body force, scales with $d_p^3$. When the particle size is small and the LHe flow speed is high, a significant streamwise displacement of the particle can occur. The strong horizontal gradient of $E(\mathbf{r})$, achieved by activating all four coils, is crucial for confining the particle within the trap region in high-Re LHe flows. The four coil configuration also provides precise control over the height of the levitation point. For instance, Figure~\ref{fig:Figure 3}b shows the contour plot of $E(\mathbf{r})$ with the current in Coil \#1  remaining the same as in Figure~\ref{fig:Figure 3}a and the currents in Coils \#2, \#3 and \#4 decrease to $N_2I_2=-900$~A, $N_3I_3=-1500$~A and $N_4I_4=-2400$~A, respectively. In this case, the height of the levitation point shifts to $y_0=0.98$~mm. Our numerical study suggests that the four-coil system enables precise adjustment of the particle levitation height from $y_0=20$~$\mu$m to $y_0=2$~mm for particles with diameters of $d_p=15-50$~$\mu$m. In all the studied cases, the maximum magnetic field strength at the levitation point is 0.014 T, which is far below the lower critical field of niobium.

\subsection*{Boundary layer study using PLV}
\subsubsection*{Study of mean velocity}
To illustrate how the PLV system can be utilized for measuring mean velocity profiles in the boundary layer, we begin by considering the equation of motion for the niobium particle, given by:
\begin{equation}
\rho_p V_p \frac{d^2 \mathbf{r}_p}{dt^2} = [-\mathbf{\nabla}E(\mathbf{r})]V_p + \mathbf{F}_d
\label{eq:motion}
\end{equation}
where $\mathbf{r}_p$ is the position vector of the particle, $V_p = \frac{\pi}{6}d_p^3$ is the particle's volume, $-\mathbf{\nabla}E(\mathbf{r})$ represents the volumetric force acting on the particle due to the combined effects of the magnetic field and gravity. The term $\mathbf{F}_d$ represents the drag force exerted by the LHe flow within the pipe. Assuming the flow velocity of LHe at the particle's location is $\mathbf{U}_f$, the drag force $\mathbf{F}_d$ is given by $\emph{\textbf{F}}_{\mathrm{d}}=-\frac{3}{4} \frac{C_{\mathrm{D}} \mathrm{Re}_{\mathrm{p}} \mu}{d_{\mathrm{p}^2}}\mathbf{V}$~\cite{melling1997tracer}, where $ \mathbf{V} = \frac{d\mathbf{r}_p}{dt}-\mathbf{U}_f $ is the relative velocity between the particle and the LHe flow. The drag coefficient $C_D$ can be calculated as~\cite{white2006viscous}:
\begin{equation}
C_D \approx \frac{24}{\mathrm{Re}_p} + \frac{6}{1 + \sqrt{\mathrm{Re}_p}} + 0.4
\label{eq:Cd}
\end{equation}
In this expression, $\mathrm{Re}_p=\rho_f\mathbf{V}d_p/\mu$ is the particle Reynolds number, where $\mu$ is the dynamic viscosity of LHe. Eq.~(\ref{eq:Cd}) applies for $\mathrm{Re}_p$ values in the range of $0\le \mathrm{Re}_p \le 2\times10^5$, which is valid for all the flows considered here.

\begin{figure}[ht]
\centering
\includegraphics[width=1\linewidth]{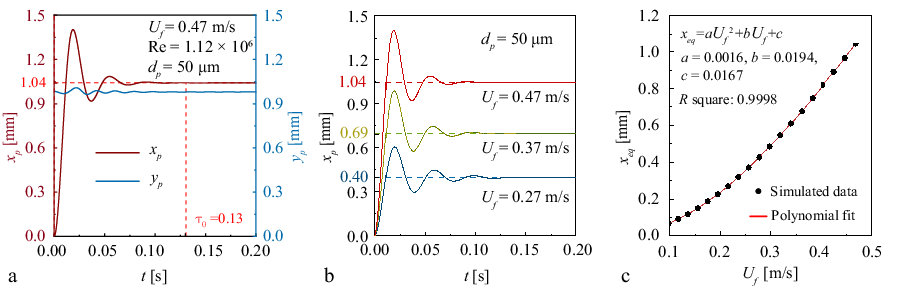}
\caption{\textbf{Position evolution of levitated particles in LHe flows.} \textbf{a} Coordinate evolution for a particle with $d_p=50$~$\mu$m, initially levitated at $y_0=0.98$~mm, with a local LHe flow velocity of $U_f=0.47$~m/s. \textbf{b} Coordinate evolution of the same particle under different flow velocities. \textbf{c} Correlation between  equilibrium position $x_{eq}$ and local flow velocity $U_f$}
\label{fig:Figure 4}
\end{figure}

Assuming a niobium particle initially held stationary at the center of the magnetic trap, Eq.~(\ref{eq:motion}) allows us to simulate its motion when a LHe flow is turned on at $t=0$. As an example, Figure~\ref{fig:Figure 4} shows the time evolution of the coordinates $x_p(t)$ and $y_p(t)$ for a particle with a diameter of $d_p=50$~$\mu$m, initially levitated at $x_p(0)=x_0=0$ and $y_p(0)=y_0=0.98$~mm. This simulation is based on a local flow velocity of $U_f=0.47$~m/s in the $x$ direction. Due to the drag force, the particle drifts downstream and exhibits a damped oscillation around the new equilibrium location at $x_{eq}=1.04$~mm. The drift in the $y$ direction is negligible, and the oscillations are significantly weaker in this direction. To characterize the damping, we introduce a damping time scale $\tau_0$, defined as the time it takes for the particle's streamwise velocity to fall below $10^{-4}$ m/s. By this time, all the oscillations around the new equilibrium location are considered negligible. For the case shown in Figure~\ref{fig:Figure 4}a, $\tau_0$ is about 0.13~s, which is much shorter than the time scale of about 10~s during which a steady flow can be maintained in the pipe by the bellows pump in a single stroke~\cite{sanavandi2022liquid}. The downstream displacement of the particle strongly depends on the flow velocity. Figure~\ref{fig:Figure 4}b shows the evolution of the streamwise coordinate $x_p(t)$ for the same particle as in Figure~\ref{fig:Figure 4}a but at three different flow velocity $U_f$. As $U_f$ increases from 0.27~m/s to 0.47~m/s, the equilibrium displacement increases from about 0.4~mm to 1.04~mm. The correlation between $x_{eq}$ and $U_f$ obtained over the flow velocity $U_f$ range from 0.1 to 0.47 m/s is shown in Figure~\ref{fig:Figure 4}c for the same particle. This correlation can be well described by a second-order polynomial expression, $x_{eq}=aU_f^2+bU_f+c$, with the coefficients $a$, $b$, and $c$ provided in the figure. Therefore, by measuring the downstream displacement $x_{eq}$, one can use this correlation to determine the corresponding flow velocity $U_f$, a key concept of PLV. The coefficients $a$, $b$, and $c$ depend on the particle size $d_p$ and the initial levitation height $y_0$, but they can be easily determined through the same analysis for different configurations.

\begin{figure}[ht]
\centering
\includegraphics[width=\linewidth]{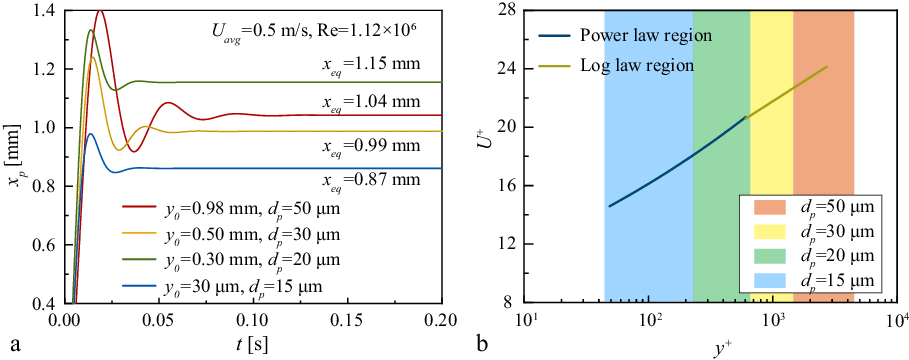}
\caption{\textbf{Boundary-layer mean velocity analysis. a} Coordinate evolution of particles with different diameters $d_p$ and levitated at varying heights $y_0$ under LHe flow with a pipe-averaged velocity of $U_{avg}=0.5$~m/s. {b} Range of wall units $y^+$ and the corresponding range of the dimensionless velocity $U^+$ that can be explored by these particles.}
\label{fig:Figure 5}
\end{figure}

To apply PLV for boundary layer velocity measurements, we can adjust the levitation height $y_0$ of the niobium particle. The local mean flow velocity varies at different heights, which results in the particle drifting different distances downstream. To illustrate this effect, consider a boundary layer described by the log law~\cite{zagarola1998mean,mckeon2004further}:
\begin{equation}
U^+ =
\begin{cases}
8.7 \, y^+{}^{0.137}, & 50 < y^+ < 600 \\
\frac{1}{0.42} \ln(y^+) + 5.6, & 600\leqslant y^+ \leqslant0.12 Ru_{\tau}/\nu.
\end{cases}
\label{eq:U+}
\end{equation}
For a particle levitated at $y_p=y_0$ from the pipe bottom surface, the corresponding wall unit is $y^+ = y_p u_{\tau}/\nu$, where the viscous velocity $u_{\tau}$ in He II pipe flows can be evaluated as $u_\tau=(\tau_w/\rho_f)^{1/2}=(\frac{1}{8}fU_{avg}^2)^{1/2}$, with $U_{avg}$ being the mean velocity averaged over the pipe cross-section, and $f$ being the measured friction factor for LHe pipe flows~\cite{swanson2000turbulent,walstrom1988turbulent}. Knowing the value of $y^+$, the dimensionless mean velocity $U^+(y^+)$ can be determined from the above log-law expression. The mean streamwise velocity at height $y_p$ in physical space is then given by $\overline{U}_f(y)=u_{\tau}U^+(y^+)$. Figure \ref{fig:Figure 5}a shows the evolution of the $x_p(t)$ coordinate for particles with diameters $d_p=15-50$ $\mu$m, levitated at different heights $y_p$ from the bottom window surface in a LHe flow with $U_{avg}=0.5$~m/s, corresponding to a pipe Reynolds number of $\mathrm{Re}=1.12\times10^6$. The results clearly show that particles of a given size drift different distances depending on their location relative to the wall, resulting in different $x_{eq}$.

In a real PLV experiment, the process works in reverse. First, a particle is levitated at height $y_0$, and then the flow is initiated. Once the particle settles at its equilibrium position, the downstream displacement $x_{eq}$ is measured. Using the previously established correlation, the mean streamwise velocity, $\overline{U}_f(y_0)$, can be determined. By varying $y_0$ and repeating the measurements, a near-wall mean velocity profile can be constructed. Nonetheless, when a particle is placed too close to the wall, the streamwise flow velocity $\overline{U}_f$ becomes too low, resulting in a downstream displacement comparable to the particle's diameter $d_p$, making it difficult to resolve. Conversely, if the particle is placed too far from the wall, where the flow velocity is much higher, the displacement may exceed the boundaries of the magnetic trap, causing the particle to lose confinement. Since the viscous drag force on the particle $F_d$ scales with $d_p^2$, while the magnetic restoring force $[-\mathbf{\nabla}E(\mathbf{r})]V_p$ scales with the particle volume $V_p$ (and therefore $d_p^3$), smaller particles tend to drift farther downstream compared to larger particles at a given flow velocity. Therefore, small particles are more suitable for probing the velocity profile near the wall, where the velocity is low. On the other hand, larger particles are better suited for exploring regions farther from the wall, where the velocity is higher and the magnetic trap is less effective at confining smaller particles. In Figure~\ref{fig:Figure 5}b, we present the calculated range of the wall unit $y^+$ that can be explored by niobium particles of four different diameters, i.e., $d_p=$15, 20, 30, and 50 $\mu$m, under the same LHe flow conditions shown in Figure~\ref{fig:Figure 5}a. By combining measurements using these different particles, we can explore the mean streamwise velocity $\overline{U}_f(y)$ over a wall unit range of $44 \leq y^+ \leq 4400$, covering the entire logarithmic law region and beyond, as illustrated in Figure~\ref{fig:Figure 5}b.

\subsubsection*{Study of turbulence intensity}
\begin{figure}[ht]
\centering
\includegraphics[width=\linewidth]{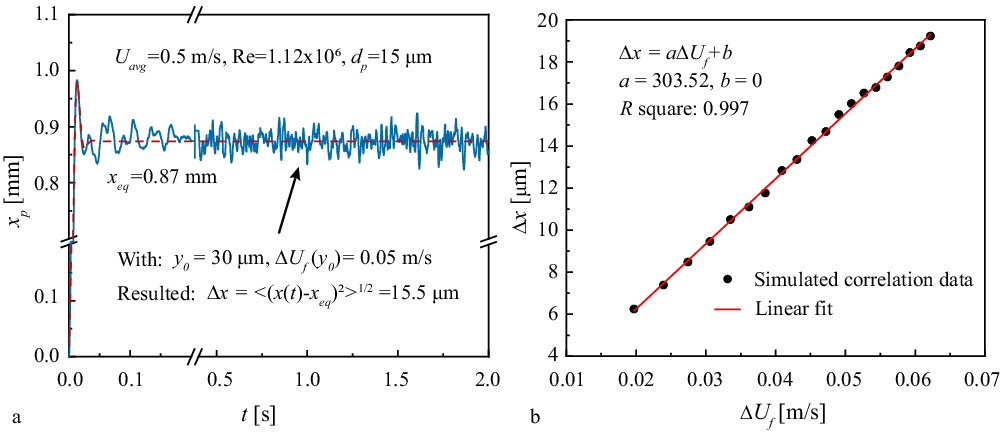}
\caption{\textbf{Boundary-layer velocity fluctuation analysis. a} Coordinate evolution of a particle with $d_p=15$~$\mu$m levitated at $y_0=30$~$\mu$m when a random velocity fluctuation with amplitude $\Delta U_f=0.05$~m/s is added to the local streamwise velocity. The pipe-averaged flow velocity is $U_{avg}=0.5$~m/s. \textbf{b} Streamwise position fluctuation amplitude $\Delta x$ as a function of $\Delta U_f$ for the particle considered in \textbf{a}.}
\label{fig:Figure 6}
\end{figure}

After the particle settles at its new downstream equilibrium position, $x_{eq}$, in the presence of a LHe flow, turbulent eddies can cause it to fluctuate around this position. The amplitude of these fluctuations is expected to correlate with the velocity fluctuations, $\Delta U_f$, in the LHe. By measuring the particle's position fluctuations, one can infer $\Delta U_f$, providing insights into the turbulence intensity profile, $\Delta U_f(y)/\overline{U}_f(y)$, within the boundary layer. To demonstrate this, we conducted simulations by introducing random velocity fluctuations, $\Delta U_f(y)$, into the mean streamwise velocity, $\overline{U}_f(y)$, in the equation of motion for the niobium particle (i.e., Eq.~(\ref{eq:motion})). Near the solid wall, turbulence is anisotropic, with streamwise velocity fluctuations dominating the other two directions\cite{jimenez2013near}. In our simulation, we adopted streamwise velocity fluctuations, $\Delta U_f(y)$, based on the near-wall turbulence intensity profile reported by M. Hultmark, \emph{et al.}\cite{PhysRevLett.108.094501}. For instance, for a niobium particle levitated at $y_0=30$~$\mu$m from the bottom window surface in the presence of a LHe flow with a pipe averaged velocity of $U_{\text{avg}}=0.5$ m/s, the streamwise velocity fluctuation $\Delta U_f$ is 0.05 m/s. Figure~\ref{fig:Figure 6}a shows the time evolution of the $x$-coordinate for a particle with diameter $d_p=15$~$\mu$m in such a flow. In this simulation, we varied the flow velocity as $U_f(t)=\overline{U}_f+\Delta U_f(t)$ at each time step $\Delta t=50$~$\mu$s, where $\Delta U_f$ follows a normal distribution with a root mean square value $\langle\Delta U_f^2(t)\rangle^{1/2}=0.05$~m/s. As seen in Figure~\ref{fig:Figure 6}a, the particle's $x$-coordinate fluctuates around the equilibrium location $x_{eq}=0.87$~mm. By analyzing the fluctuations in the time window from 0.1 s to 20~s, we obtain the mean fluctuation amplitude to be $\Delta x =\langle (x-x_{eq})^2 \rangle^{1/2}=15.5$~$\mu$m. For practical PLV application, a correlation between $\Delta x$ and $\Delta U_f$ is needed. To obtain this correlation, we repeated the simulation by varying $\Delta U_f$ in the range of 0.02~m/s to 0.065~m/s. The results are shown in Figure \ref{fig:Figure 6}b, which reveal a strong linear correlation between $\Delta x$ and $\Delta U_f$ as $\Delta x=a\Delta U_f$, with the coefficient $a$ provided in the figure. Again, the coefficient $a$ depends on the particle size $d_p$ and the initial levitation height $y_0$. For particles with different sizes and levitation heights, the correlation between $\Delta x$ and $\Delta U_f$ can be determined through similar numerical analysis. By experimentally measuring the fluctuations in particle motion at various levitation heights and applying the corresponding correlation tailored to each height, one can gain a comprehensive understanding of the turbulence intensity profile within the boundary layer.

\section*{Discussion}
We would like to point out that fluid shear within the boundary layer can induce rotation in a levitated spherical particle. This rotation results in a lift force due to the Magnus effect and can also cause a deviation in the drag coefficient $C_D$ from the standard expression given in Eq.~(\ref{eq:Cd}). These effects become significant for large particles and in regions with high velocity gradients. However, for the micron-sized particles considered in our proposed PLV system, we estimate that these deviations are minimal. Based on the correlations reported by Ryoichi \emph{et al.}~\cite{kurose1999drag}, we calculate that both the deviation in $C_D$ and the ratio of lift force to particle gravity should remain below a few percent for all particle sizes and flow conditions used in our calculations. To eliminate the effects of particle rotation due to flow shear, one could consider fabricating the particles to be slightly non-spherical. Non-spherical particles tend to orient in a specific direction within the magnetic trap, effectively preventing undesired rotation caused by the flow. This design strategy may help improving measurement accuracy.

We would also like to highlight that multiple particles with different diameters can be placed in the storage pit from the beginning. Additionally, by coating Teflon or other plastic beads with niobium, it is possible to produce particles with varying overall densities. This introduces an interesting dynamic: particles of the same density but different sizes will be levitated at nearly the same height but displaced to different downstream locations in the presence of LHe flow. Conversely, particles of the same size but different densities would experience similar downstream displacements but will be levitated at different heights. By introducing multiple particles and simultaneously measuring their motion, one can capture both temporal and spatial correlations of velocity fluctuations within the boundary layer. This multi-particle approach offers a unique opportunity to study turbulence structures and correlations in the thin boundary layer, providing a level of detail that is otherwise impractical to achieve. In summary, the combination of the PLV system and the distinctive properties of LHe in a cryogenic environment opens up exciting new possibilities, fully unlocking the potential of LHe in high Reynolds number turbulence research.

\section*{Methods}
\subsection*{Magnetic field generated by the coils}
Consider a single solenoid, which is a cylindrical coil of length $2L$ and radius $R$, tightly wound with closely packed current loops. The solenoid is centered at the origin of a cylindrical coordinate system $(\rho, \phi, z)$, with its axis aligned along the $z$-axis. Let $I$ represent the current in each loop, and $N$ denote the total number of loops. The magnetic field generated by this solenoid in three-dimensional space can be calculated using the Biot-Savart law \cite{jackson1999curious}. In terms of elliptic integrals~\cite{caciagli2018exact,derby2010cylindrical}, the magnetic field can be expressed as:
%\begin{equation}
%B_z = \frac{\mu_0 NI}{2} \left( \frac{z + L}{\sqrt{(z + L)^2 + R^2}} - \frac{z - L}{\sqrt{(z - L)^2 + R^2}} \right)
%\label{eq:Bz}
%\end{equation}
%For general case, the magnetic field can be calculated as:
\begin{equation}
B_\rho = \frac{\mu_0 NI R}{\pi} \left[ \alpha_+ P_1(k_+) - \alpha_- P_1(k_-) \right]
\label{eq:B_rou}
\end{equation}
\begin{equation}
B_z = \frac{\mu_0 NI R}{\pi (\rho + R)} \left[ \beta_+ P_2(k_+) - \beta_- P_2(k_-) \right].
\label{eq:Bz2}
\end{equation}
Due to  axial symmetry, $B_\phi$ component is absent. Functions $P_1$ and $P_2$ are defined as:
\begin{equation}
P_1(k) = K(k) - \frac{2}{k^2} \left[ K(k) - E(k) \right]
\label{eq:P_1(k)}
\end{equation}
\begin{equation}
P_2(k) = -\frac{\gamma}{1 - \gamma^2} \left[ \Pi(1 - \gamma^2, k) - K(k) \right] - \frac{1}{1 - \gamma^2} \left[ \gamma^2 \Pi(1 - \gamma^2, k) - K(k) \right]
\label{eq:P_2(k)}
\end{equation}
where
\begin{equation}
\alpha_\pm = \frac{1}{\sqrt{\xi_\pm^2 + (\rho + R)^2}} \quad \quad
\beta_\pm = \xi_\pm \alpha_\pm \quad \quad
\xi_\pm = z \pm L\quad \quad
\gamma = \frac{\rho - R}{\rho + R}\quad \quad
k_\pm^2 = \frac{4 \rho R}{\xi_\pm^2 + (\rho + R)^2}
\label{eq:subs}
\end{equation}
and the functions $K(k)$, $E(k)$, and $\Pi(1 - \gamma^2, k)$ are the complete elliptic integrals of the first, second, and third kind:
\begin{equation}
\left\{
\begin{aligned}
K(k) &= \int_0^{\frac{\pi}{2}} \frac{d\theta}{\sqrt{1 - k^2 \sin^2 \theta}} \\
E(k) &= \int_0^{\frac{\pi}{2}} \sqrt{1 - k^2 \sin^2 \theta} \, d\theta \\
\Pi(1 - \gamma^2, k) &= \int_0^{\frac{\pi}{2}} \frac{d\theta}{\left[ 1 - (1 - \gamma^2) \sin^2 \theta \right] \sqrt{1 - k^2 \sin^2 \theta}}
\end{aligned}
\right.
\label{eq:elliptic integral}
\end{equation}

In our analysis, we calculate the magnetic field produced by each superconducting coil $\mathbf{B}_i(\mathbf{r})$ with $i=1,2,3,4$. The total magnetic field $\mathbf{B}(\mathbf{r})$ is the sum of the contributions from all individual coils $\mathbf{B}(\mathbf{r})=\sum_{i=1}^4\mathbf{B}_i(\mathbf{r})$.

\subsection*{Key parameter specifications for numerical analysis}
In our study, we used MATLAB to perform several key calculations: 1) the magnetic field, $\mathbf{B}(\mathbf{r})$, generated by the coils; 2) contour plots of the volumetric potential energy $E(\mathbf{r})$ for superconducting niobium particles in the magnetic field; and 3) the motion of levitated particles under various LHe flows in the pipe. Given the axial symmetry of the coil system, we employed a computational domain of $16\times12$~mm$^2$ in the $x$-$y$ plane, with the $x$-axis pointing in the flow direction and the $y$-axis aligned along the coils' axial direction. This domain was discretized using a square grid with a spatial resolution of 10$\mu$m in both directions. For the particle motion calculations, a time step of 50~$\mu$s was used. The chosen grid size and time step were validated by confirming that the results were independent of both parameters, ensuring convergence. The particle was assumed to be pure niobium, with a density of 8570 kg/m$^3$, a critical temperature of 9.2~K, and a lower critical magnetic field of 0.15~T at 4.2~K. The LHe was modeled with a density of 145 kg/m$^3$ and a kinematic viscosity of $8.929\times10^{-9}$m$^2$/s\cite{Donnelly-1998-JPCRD}. The friction factor for LHe in high Reynolds number pipe flows has been reported in the literature\cite{swanson2000turbulent,walstrom1988turbulent}. For instance, for a LHe pipe flow with $\mathrm{Re}=1.12\times10^6$ considered in our analysis, the friction factor is $f=0.0124$. The particle’s levitation point was determined by locating the minimum energy point within the trapping region.

\bibliography{Refs}

\section*{Acknowledgements}
Y.Q. and W.G. acknowledge support from the Gordon and Betty Moore Foundation through Grant DOI 10.37807/gbmf11567. This work was conducted at the National High Magnetic Field Laboratory at Florida State University, supported by the National Science Foundation Cooperative Agreement No. DMR-2128556 and the state of Florida. The authors also wish to thank Lauren Roche for her assistance in the design of the superconducting coils.

\section*{Author contributions statement}
W.G. designed the research. Y.Q. designed the coil configuration and conducted the numerical simulations. Y.Q. and W.G. analyzed the results and collaboratively wrote and reviewed the manuscript.

\section*{Data availability statement}
The datasets used and/or analysed during the current study available from the corresponding author on reasonable request.

\section*{Additional information}
\textbf{Competing interests}: The authors declare no competing interests.

\end{document}